\documentclass[
    ,final            
  ]
  {aipproc}

\usepackage{epsfig,epsf}
\usepackage{amssymb,amsbsy,amsmath,amsfonts}
\layoutstyle{8x11single}

\def\beq{\begin{equation}}
\def\eeq{\end{equation}}
\def\bea{\begin{eqnarray}}
\def\eea{\end{eqnarray}}
\def\beqa{\begin{equation}\begin{array}{l}}
\def\eeqa{\end{array}\end{equation}}
\def\eqlab#1{\label{eq:#1}}
\def\figlab#1{\label{fig:#1}}
\def\tablab#1{\label{tab:#1}}

\def\barr{\left(\begin{array}{c}}
\def\earr{\end{array}\right)}
\def\bmat{\left(\begin{array}{cc}}
\def\emat{\end{array}\right)}

\def\Eqref#1{Eq.~(\ref{eq:#1})}

\def\Figref#1{Fig.~\ref{fig:#1}}

\def\Tabref#1{Table \ref{tab:#1}}

\def\sla#1{#1  \!\!\!\!\slash}

\def\al{\alpha}

\def\ga{\gamma} 
\def\de{\delta} \def\De{\Delta}
\def\vDe{\varDelta}
  \def\eps{\epsilon}

\def\la{\lambda} \def\La{{\Lambda}}

\def\si{\sigma}

\def\pa{\partial}

\def\pa{\partial}

\def\nn{\nonumber}

\def\lag{{\mathcal L}}

\def\mathscr{\mathcal}
\def\N{N}

\def\3d{3-D}

\def\ol#1{\overline{#1}}

\def\ceft{$\chi$EFT}

\begin{document}

\title{The $\ga N\to \De$ transition in chiral effective-field theory}


\author{Vladimir Pascalutsa}{
  address={Physics Department, The College of William \& Mary,
Williamsburg, VA 23187, USA}
,altaddress={Theory Center, Jefferson Lab, 12000 Jefferson Ave, Newport News,
VA 23606, USA}
}

\author{Marc Vanderhaeghen}{
  address={Physics Department, The College of William \& Mary,
Williamsburg, VA 23187, USA}
,altaddress={Theory Center, Jefferson Lab, 12000 Jefferson Ave, Newport News,
VA 23606, USA}
}

\begin{abstract}
We describe the pion electroproduction
processes in the $\Delta$(1232)-resonance region within the framework of
chiral effective-field theory. By studying the
observables of pion electroproduction in a next-to-leading order
calculation we are able to make
predictions and draw conclusions on the properties of the $N\to \Delta$
electromagnetic form factors.
\end{abstract}

\maketitle


\indent

For the general overview and motivation of the electromagnetic 
$N\to \De$ transition we refer the reader to the opening 
part of these Proceedings and to our recent review~\cite{Pascalutsa:2006up}.
In this short contribution we will focus on the chiral 
effective-field theoretic study of the $N\to \De$ transition.
 
\section{Not-so-low energy expansion}

The strength of chiral interactions 
goes with derivatives of pion fields which allows one to organize
a perturbative expansion in powers of pion momentum and mass  --- 
the chiral perturbation theory ($\chi$PT)~\cite{Weinberg:1978kz}.
The small expansion parameter is $p/\La_{\chi SB}$, where $p$
is the momentum and $\La_{\chi SB}\sim 4\pi f_\pi \approx 1$ GeV
stands for the scale of spontaneous chiral symmetry breaking. 
Based on this expansion
one should be able to systematically compute
the pion-mass dependence of static quantities, such as nucleon mass,
magnetic moments, as well as the momentum dependence of scattering
processes, such as pion-pion and pion-nucleon scattering. 
This is an effective-field theory (EFT) expansion, in this
case a low-energy expansion of QCD. One expects to obtain exactly 
the same answers as from QCD directly, provided the low-energy
constants (LECs) --- the parameters of the effective Lagrangian --- 
are known, either from experiment or by matching to QCD itself
({\it e.g.}, by fitting to the lattice QCD results). 


One of the principal ingredients of an EFT expansion is 
{\it power counting}. The power counting assigns an order
to Feynman graphs arising in loopwise expansion of the amplitudes,
and thus defines which graphs need to be computed at a given order in the
expansion. In a way it simply is a tool to estimate the size of
different contributions without doing explicit calculations.
The main requirement on a power-counting scheme
is that it should estimate the relative size of various contributions
correctly. 
In $\chi$PT with pions and nucleons alone~\cite{GSS89}, the power counting
for a graph with $L$ loops, $N_\pi$ ($N_N$) internal pion (nucleon)
lines, and $V_k$ vertices from $k$th-order Lagrangian, estimates
its contribution to go as $p^{n_{\chi \mathrm{PT}}}$, with the power
given by:
\beq
\eqlab{chptindex}
n_{\chi \mathrm{PT}} =  4 L  - 2 N_\pi - N_N + \sum_k k V_k\,.
\eeq
Note that in the manifestly Lorentz-covariant formalism 
this power counting holds only in a specific renormalization
scheme~\cite{Gegelia:1999gf,Pascalutsa:2005nd}.

A cornerstone principle of effective field theories in general is
{\it naturalness}, meaning that the dimensionless
LECs must be of natural size, {\it i.e.}, of order of unity.
Any significant fine-tuning of even a single LEC leads, obviously,
to a break-down of the EFT expansion. Therefore, even if an EFT
describes the experimental data, but at the expense of fine-tuned
LECs, the EFT expansion fails. Such an EFT can still be useful for getting insights
into the physics beyond the EFT itself. Namely, by looking
at the form of the fine-tuned operators, one might be able to 
deduce which contributions are missing.  

For instance, it is  known that the NLO 
$\chi$PT description of the pion-nucleon elastic scattering,
near threshold, requires relatively large values for some of the LECs.
It is not difficult to see that the operators corresponding 
with those unnatural LECs can be matched to the ``integrated out''
$\De$-resonance contributions. 
The problem is  that the $\De$ is relatively light, its excitation
energy, $\vDe\equiv M_\De-M_N\sim 0.3$ GeV, is still quite
small compared to $\La_{\chi SB} \sim 1$ GeV.
Integrating out the $\De$-isobar degrees of freedom
corresponds to an expansion in powers of $p/\vDe$, with $p\sim m_\pi$,
which certainly is not as good of an expansion as the one 
in the meson sector, in powers of $p/\La_{\chi SB}$.
\newline
\indent
The fine-tuning of the ``Deltaless'' $\chi$PT seems to be 
lifted by the inclusion of an explicit $\De$-isobar.
Also, the limit of applicability of the EFT expansion
is then extended to momenta of order of the resonance excitation
energy, $p\sim \vDe$. Such momenta can still be considered as soft,
as long as $\vDe /\La_{\chi SB}$ can be treated as small.  
The resulting $\chi$PT with pion, nucleon, and $\De$-isobar degrees
of freedom has two distinct light scales: $m_\pi$ and $\De$.
Perhaps the most straightforward way to proceed is to organize a
simultaneous expansion in two different small parameters:
$$\eps = m_\pi /\La_{\chi SB} \,\,\,\, \mbox{and} 
\,\,\,\, \de = \vDe /\La_{\chi SB}.$$
However, for power counting purposes it is certainly more convenient
to have a single small parameter and thus a relation between
$\eps$ and $\de$ is usually imposed. We stress that the relation
is established only at the level of power counting and not in the 
actual calculations of graphs. 
At present two such relations between $\eps$ and $\de$
 are used in the literature: 
\begin{itemize} 
\item{} $\eps \sim \de$, 
the so-called ``Small Scale Expansion''~\cite{JeM91a,HHK97}, here refered to as
the ``$\eps$-expansion'';  \\
\item{}  $\eps \sim \de^2$, the 
``$\de$-expansion''~\cite{Pascalutsa:2002pi}.
\end{itemize}
The table below (\Tabref{compare}) summarizes the counting of momenta in the
three expansions: Deltaless  ($\sla{\De}$-$\chi$PT), 
$\eps$-expansion, and $\de$-expansion.
\begin{table}[h]
\begin{tabular}{|c||c|c|}
\hline
EFT &  $\quad p\sim m_\pi \quad$ & $\quad p\sim \Delta \quad$\\
\hline
$\sla{\De}$-$\chi$PT & ${\mathcal O}(p)$        & ${\mathcal O}(1)$\\
 $\eps$-expansion  & ${\mathcal O}(\epsilon) $ & ${\mathcal O}(\epsilon) $\\
$\delta$-expansion & ${\mathcal O}(\delta^2) $ & ${\mathcal O}(\delta) $\\
\hline
\end{tabular} 
\caption{The counting of momenta in 
the three different $\chi$EFT expansions.}
\tablab{compare}
\end{table}
\newline
\indent
An unsatisfactory feature of the $\eps$-expansion is 
that the $\De$-resonance 
contributions are always estimated to be of the
same size as the nucleon contributions. 
In reality (revealed by actually computing these contributions),
they are {\it suppressed} at low energies and {\it dominate} 
in the the $\De$-resonance region. 
Thus, apparently the power-counting in the $\eps$-expansion
{\it overestimates} the $\De$-contributions
at lower energies and {\it underestimates} them
at the resonance energies. 
The $\de$-expansion improves on this aspect, as is briefly described
in what follows.
\newline
\indent
In the $\de$-expansion, the power counting depends
on the energy domain, 
since  in the {\it low-energy region} ($p\sim m_\pi$)
and the {\it resonance region} ($p\sim \vDe$), the momentum
counts differently, see \Tabref{compare}. This dependence 
most significantly affects the power counting of the direct resonance
exchanges --- the 
one-Delta-reducible (ODR) graphs. Figure~\ref{fig:ODR}
illustrates examples of the ODR graphs for the case of Compton scattering
on the nucleon. These graphs are all characterized by having
a number of ODR propagators, each going as
\beq
S_{ODR}\sim \frac{1}{s-M_\De^2} \sim \frac{1}{2M_\De}\frac{1}{p-\vDe}\, ,
\eeq
where $p$ is the soft momentum, in this case given by the photon energy.
In contrast the nucleon propagator in analogous graphs would go
simply as $S_N\sim 1/p$.
\begin{figure}[t,b]
\centerline{  \epsfxsize=10cm
  \epsffile{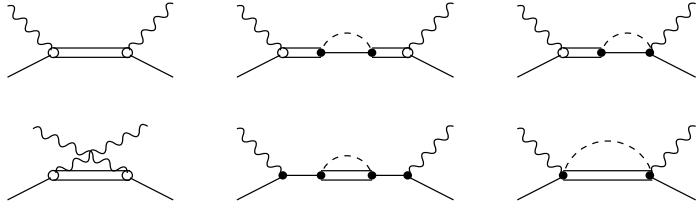} 
}
\caption{Examples of the one-Delta-reducible (1st row) and
 the one-Delta-irreducible (2nd row) graphs in Compton scattering. }
\figlab{ODR}
\end{figure}
\newline
\indent
Therefore, in the low-energy region, the $\De$ and nucleon 
propagators would
count respectively as ${\mathcal O}(1/\de)$ and ${\mathcal O}(1/\de^2)$, 
the $\De$ being suppressed by one power of the small parameter
as compared to the nucleon. 
In the resonance region, the ODR graphs obviously 
all become large. Fortunately they all can be subsumed, leading
to ``dressed'' ODR graphs with a definite power-counting index.
Namely, it is not difficult to see that the resummation of
the classes of ODR graphs results 
in ODR graphs with only a single ODR propagator of
the form
\beq
S_{ODR}^\ast = \frac{1}{S_{ODR}^{-1} - \Sigma }
\sim \frac{1}{p-\vDe-\Sigma}\,,
\eeq
where $\Sigma$ is the $\De$ self-energy.
The expansion of the self-energy begins with $p^3$, and hence
in the low-energy region 
does not affect the counting of the $\De$ contributions. However,
in the resonance region the self-energy not only ameliorates
the divergence of the ODR propagator at $s=M_\De^2$ but also
determines power-counting index of the propagator.
Defining the $\De$-resonance region formally as the region of $p$
where
\beq
|p-\vDe | \leq \de^3 \La_{\chi SB}\,,
\eeq
we deduce that an ODR propagator, in this region, counts
as ${\mathcal O}(1/\de^3)$. Note that the nucleon propagator in
this region counts as ${\mathcal O}(1/\de)$, hence is
suppressed by two powers as compared to ODR propagators.
Thus, within the power-counting scheme we have the mechanism for
estimating correctly  the relative size
of the nucleon and $\De$ contributions in the two energy domains.
In \Tabref{counting} we summarize the counting of the nucleon,
ODR, and one-Delta-irreducible (ODI) propagators in 
both the $\eps$- and $\de$-expansion.
\begin{table}[htb]
{\centering \begin{tabular}{||c|c||c|c||}
\hline
&  $\eps$-expansion & 
\multicolumn{2}{|c||}{$\de$-expansion} \\
\cline{2-4} 
  & $p/\La_{\chi SB}\sim \eps$  &  $p\sim m_\pi$ & $p\sim \vDe$ \\
\hline
$\,S_N\,$ &  $1/\eps$ & $1/\de^2$ & $1/\de$\\
$\,S_{ODR}\,$ & $1/\eps$ & $1/\de$ & $1/\de^3$\\
$\,S_{ODI}\,$ & $1/\eps$ & $1/\de$ & $1/\de $ \\
\hline
\end{tabular} \par }
\caption{The counting for the nucleon, one-Delta-reducible (ODR), and
one-Delta-irreducible (ODI) propagators in the two different expansion
schemes. The counting in the $\de$-expansion depends on the energy domain.}
\tablab{counting}
\end{table}
\newline
\indent
We conclude this discussion by giving the general
formula for the power-counting index in the $\de$-expansion.
The power-counting index, $n$, of a given graph
simply tells us that the graph is of the size of ${\mathcal O}(\de^n)$.
For a graph with $L$ loops, $V_k$ vertices of
dimension $k$, $N_\pi$ pion propagators, $N_N$
nucleon propagators, $N_\De$ Delta propagators, $N_{ODR}$
ODR propagators and  $N_{ODI}$ ODI propagators 
(such that $N_\De=N_{ODR}+N_{ODI}$) the index is
\beq
\eqlab{PCindex}
n = \left\{ \begin{array}{cc} 2 n_{\chi \mathrm{PT}} - N_\De\,, 
& p\sim m_\pi ; \nn\\
n_{\chi \mathrm{PT}} - 3N_{ODR} - N_{ODI}\,, & p\sim \De, \end{array}\right.
\eeq
where $ n_{\chi \mathrm{PT}}$, given by \Eqref{chptindex}, 
is the index of the graph in $\chi$PT with no $\De$'s. For further details
on the $\de$ counting we refer to Ref.~\cite{Pascalutsa:2002pi}. 

The $\ga N\De$ form factors have been examined 
in both the $\eps$-expansion \cite{Gellas:1998wx,Gail}
 and the $\delta$-expansion \cite{Pascalutsa:2005ts,Pascalutsa:2005vq}. 
In the following
we focus on the latter analysis.

\section{Pion electroproduction to next-to-leading order}

The $N \to \De$ transition can be induced by a pion or a photon.
The correponding effective Lagrangians are written as:
\begin{subequations}
\eqlab{lagran}
\bea
\lag^{(1)}_{N\De} &=&  \frac{i h_A}{2 f_\pi M_\De}
\ol N\, T^a \,\ga^{\mu\nu\la}\, (\pa_\mu \De_\nu)\, \pa_\la \pi^a 
+ \mbox{H.c.}, \\
\lag^{(2)}_{N\De} &=&   \frac{3 i e g_M}{2M_N (M_N + M_\Delta)}\,\ol N\, T^3
\,(\pa_{\mu}\De_\nu) \, \tilde F^{\mu\nu}  + \mbox{H.c.},\\
\lag^{(3)}_{N\De} &=&  \frac{-3 e}{2M_N (M_N + M_\Delta)} \ol N \, T^3
\ga_5 \left[ g_E (\pa_{\mu}\De_\nu) 
+  \frac{i g_C}{M_\De} \ga^\al  
(\pa_{\al}\De_\nu-\pa_\nu\De_\al) \,\pa_\mu\right] F^{\mu\nu}+ \mbox{H.c.},
\;\;\;\;\; 
\eea
\end{subequations}
where $N$, $\De_\mu$, $\pi$ stand respectively for the
nucleon (spinor, isodublet), $\Delta$-isobar (vector-spinor, isoquartet),
pion (pseudoscalar, isovector) fields;  $F^{\mu\nu}$ and $\tilde F^{\mu\nu}$
are the electromagnetic field strength and its dual,
$T^a$ are the isospin-1/2-to-3/2 transition $(2\times 4$) matrices.
The coupling constants $h_A$, $g_M$, $g_E$, and $g_C$ are 
the LECs describing the $N\to\De$ tansition at the tree level.

The $\ga N\to \De$ transition is traditionally studied
in the process of  pion electroproduction on the nucleon.
Let us consider this process to NLO in the $\de$ expansion.
Since
we are using the one-photon-exchange approximation,\footnote{For first analyses
of the two-photon-exchange effects in the $\ga N\to \De$
transition see Refs.~\cite{Pascalutsa:2005es,Kondratyuk:2006ig}.} 
the pion photoproduction can be viewed as
the particular case of electroproduction at $Q^2=0$.
\begin{figure}
\centerline{  \epsfxsize=11cm
  \epsffile{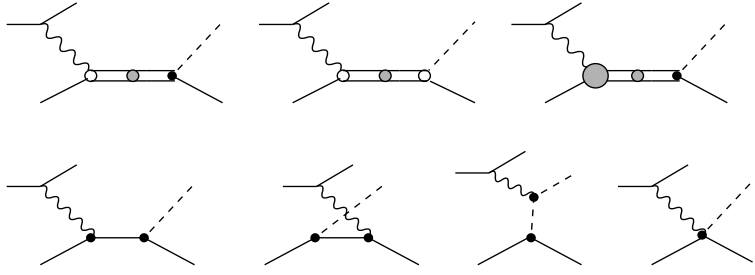} 
}
\caption{Diagrams for the $e N \to e \pi N $ reaction 
to LO and  NLO in the $\delta$-expansion. The dots denote
the vertices from the 1st-order Lagrangian, while the
circles are the vertices from the 2nd order Lagrangian ({\it e.g.}, the 
$\ga N\De$-vertex in the first two graphs is the $g_M$ coupling from $\lag^{(2)}$).}
\figlab{diagrams}
\end{figure}
\begin{figure}
\centerline{  \epsfxsize=11cm
  \epsffile{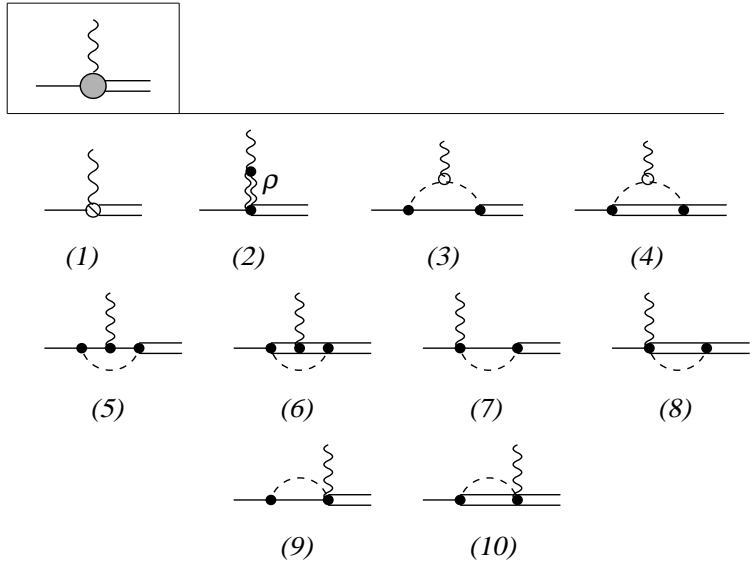} 
}
\caption{The $\ga N\De$ vertex at ${\mathcal O}(\de^3)$.
The sliced vertex (1) stands for the $g_E$ and $g_C$ couplings
from the 3rd order Lagrangian. The wiggly double line in (2) stands
for the vector-meson propagator.}
\figlab{NLOvertex}
\end{figure}
The pion electroproduction amplitude to NLO in the $\de$-expansion, in the
resonance region, is given by the graphs in  \Figref{diagrams}, 
where the shaded
blob in the 3rd graph denotes the NLO $\ga N\De$ vertex, given by the graphs
in \Figref{NLOvertex}. The 1st graph in \Figref{diagrams} enters at the LO,
which here is ${\mathcal O}(\de^{-1})$. 
All the other graphs in \Figref{diagrams}
are of NLO$={\mathcal O}(\de^{0})$. Note that the $\De$-resonance contribution
at NLO is obtained by going to NLO in either the $\pi N\De$ vertex (2nd graph)
or the $\ga N\De$ vertex (3rd graph). Accordingly, 
the $\De$ self-energy in these graphs is included, respectively,
to NLO. 
The vector-meson diagram,  \Figref{NLOvertex}(2), contributes to 
NLO for $Q^2\sim \La\De$. One includes
it effectively by giving the $g_M$-term a dipole $Q^2$-dependence 
(in analogy to how it is usually done
for the nucleon isovector form factor): 
\beq
g_M\to \frac{g_M}{(1+Q^2/0.71\,\mbox{GeV}^2)^{2}}. 
\eeq
The analogous
effect for the $g_E$ and $g_C$ couplings begins at N$^2$LO. 
\newline
\indent
An important observation is that at $Q^2=0$ 
only the imaginary part (unitarity cut) of the loop graphs in 
\Figref{NLOvertex}
contributes to the NLO amplitude. Their real-part contributions, after the
renormalization of the LECs, begin to contribute at N$^2$LO, for 
$Q^2\ll \vDe\La_{\chi SB}$.
At present we will consider only the NLO calculation where the 
$\pi\De$-loop contributions to the $\ga N\De$-vertex are omitted 
since they do not give the imaginary contributions
in the $\De$-resonance region. We emphasize that such loops might 
become important at this order for $Q^2\sim \vDe\La_{\chi SB}\sim 0.3$ GeV$^2$ 
and should be included for the complete NLO 
result. The present calculation is thus 
restricted to values $Q^2 < 0.3$ GeV$^2$.

\begin{figure}[b]
\centerline{
\includegraphics[width=0.32\columnwidth]{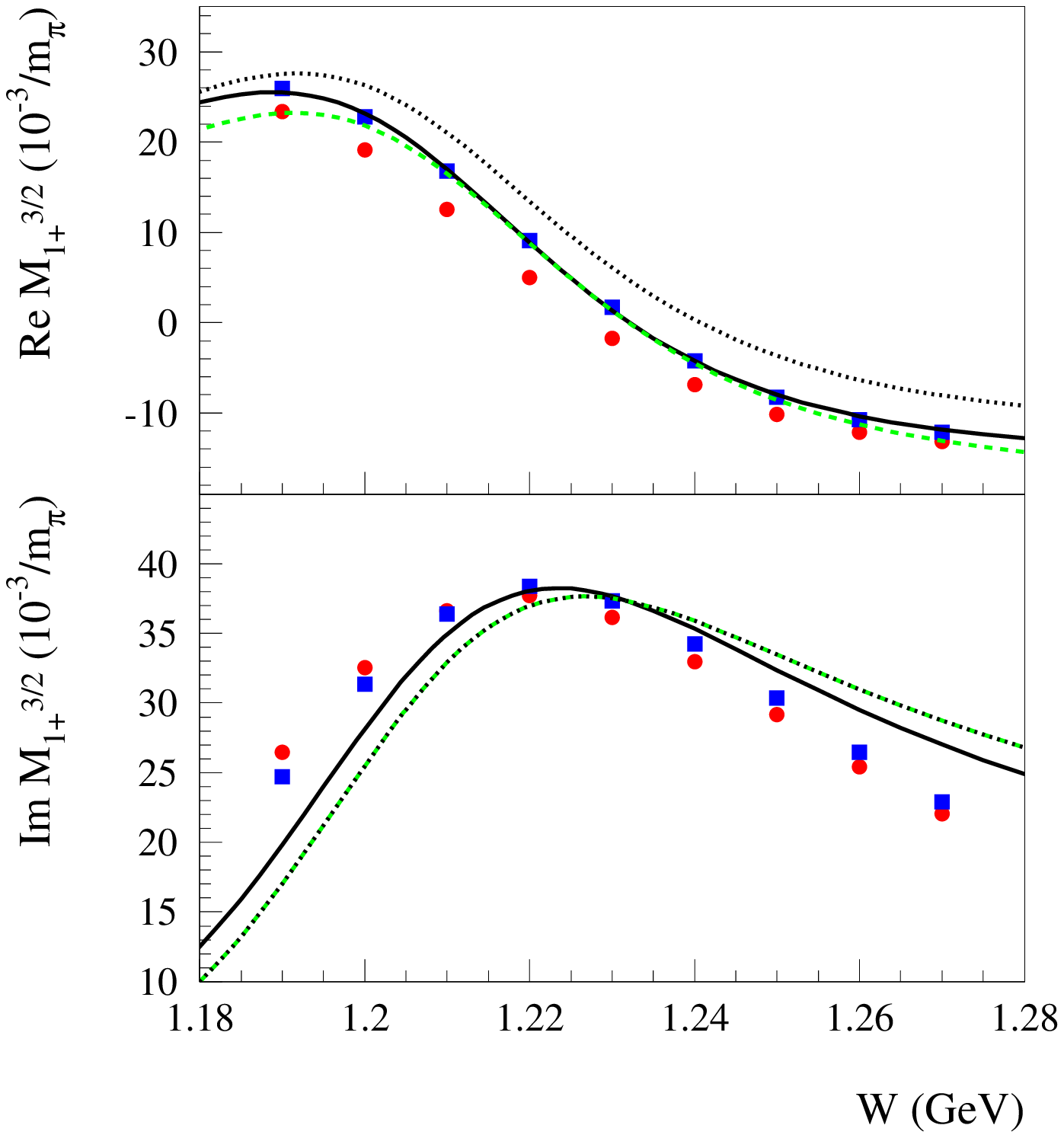}
\includegraphics[width=0.32\columnwidth]{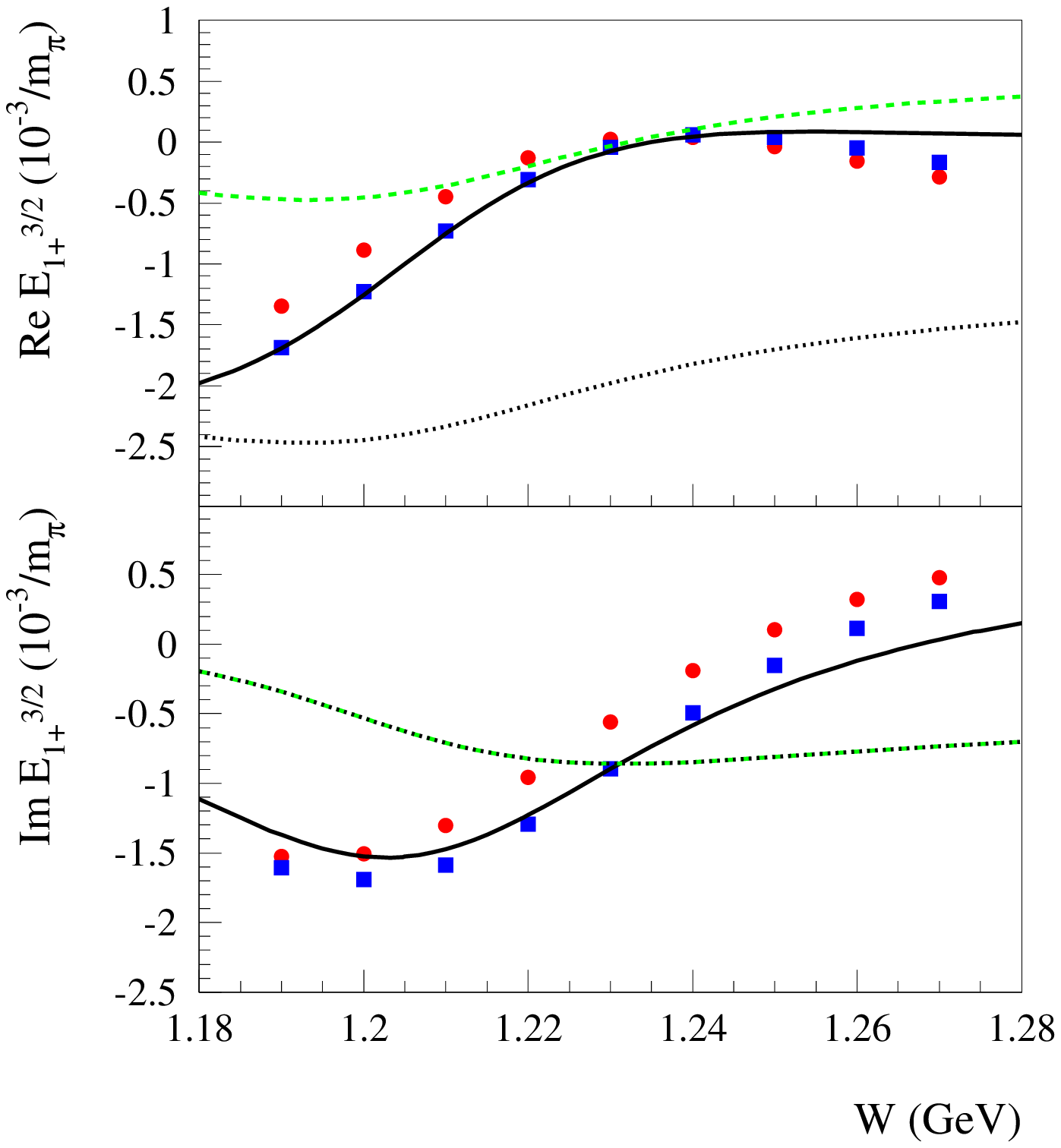}
\includegraphics[width=0.32\columnwidth]{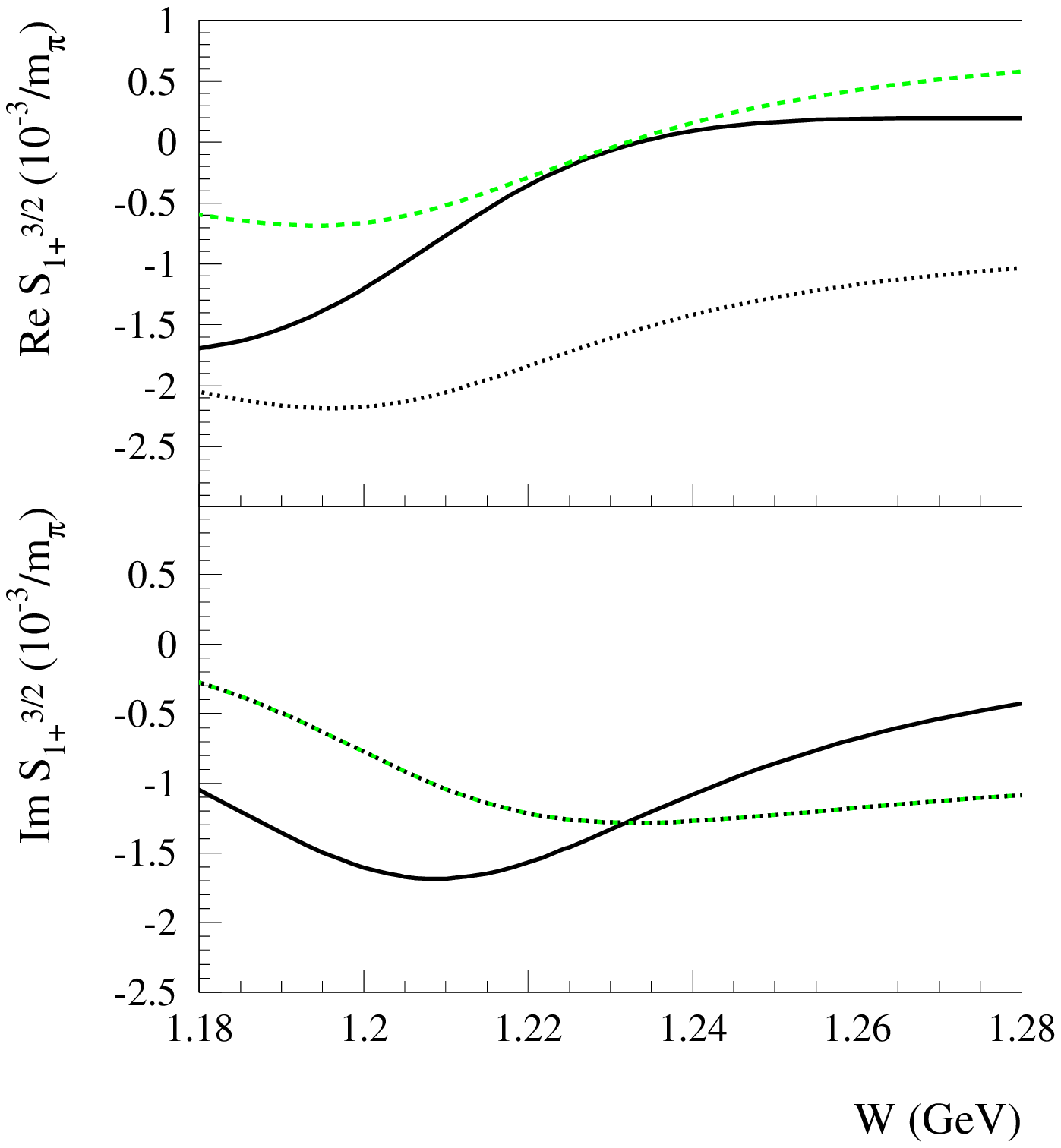}
}
\caption{
(Color online) The 
$M_{1+}^{(3/2)}$, $E_{1+}^{(3/2)}$, and $S_{1+}^{(3/2)}$ multipoles
for pion photoproduction as function of the invariant energy.  
Green dashed curves: 
$\Delta$ contribution without the $\ga N\De$-vertex loop corrections,
[i.e., only the first three graphs in \Figref{diagrams} with
\Figref{NLOvertex}(1) contribution are taken into account].  
Blue dotted curves: adding the Born contributions, 2nd line in 
\Figref{diagrams}, to the dashed curves. 
Black solid curves: the NLO calculation, includes 
all graphs in \Figref{diagrams} as well as the loop corrections. 
The data point are from the 
SAID analysis~(FA04K)~\protect\cite{GWU} (red circles), and from the 
MAID 2003 analysis~\protect\cite{MAID} (blue squares).
}
\figlab{gap_pin_m1mult}
\end{figure}
In Fig.~\ref{fig:gap_pin_m1mult}
shown are the resonant  multipoles 
$M_{1+}^{(3/2)}$, $E_{1+}^{(3/2)}$, and $S_{1+}^{(3/2)}$ as function of the invariant energy $W=\sqrt{s}$
around the resonance position and $Q^2=0$. 
The $M_{1+}^{(3/2)}$ and $E_{1+}^{(3/2)}$ 
multipoles are well established by the SAID~\cite{GWU} 
and MAID~\cite{MAID} empirical partial-wave solutions, 
thus allowing one to fit 
two of the three $\ga N\De$ LECs at this order as:~$g_M = 2.97$, $g_E = -1.0$.
The third LEC is adjusted to for a best description of the
pion electroproduction data at low $Q^2$ (see ), yielding
 $g_C=-2.6$.
The latter values translate into $G_M^\ast=3.04$, $G_E^\ast=0.07$, and $G_C^\ast=1.00$
for the Jones--Scadron form-factors~\cite{Jones:1972ky} at $Q^2=0$.    
As is seen from the figure,
the NLO results (solid curves) give a good description of the energy 
dependence of the resonant multipoles in 
a window of 100 MeV around the $\Delta$-resonance position.
These values yield $R_{EM}= -2.2$ \% and $R_{SM}= -3.4$ \%.

The dashed curves in these figures
show the contribution of the $\Delta$-resonant diagram of \Figref{diagrams}
{\it without} the NLO loop corrections in \Figref{NLOvertex}.
For the $M_{1+}$ multipole this is the LO and part of the NLO contributions.
For the $E_{1+}$ and $S_{1+}$ multipole
the LO contribution is absent (recall that $g_E$ and $g_C$ coupling
are of one order higher than the  $g_M$ coupling). 
Hence,  the dashed curve represents
a partial NLO contribution to $E_{1+}$ and $S_{1+}$.

Note that such a purely resonant contribution without the loop corrections 
satisfies unitarity in the sense of the Fermi-Watson 
theorem, which states that the phase of a pion 
electroproduction amplitude ${\mathcal M}_l$ is given by the
corresponding pion-nucleon phase-shift:
${\mathcal M}_l = | {\mathcal M}_l | \, \exp({i\de_l})$.
As a direct consequence of this theorem, 
 the real-part of the resonant multipoles must vanish at the resonance 
position, where the phase-shift crosses $90$ degrees.

Upon adding the non-resonant  Born graphs (2nd line in \Figref{diagrams}) 
to the dashed curves, one obtains the dotted curves. 
The non-resonant contributions are purely real at this order and hence 
the imaginary part of the multipoles do not change.
While this is consistent with unitarity for the non-resonant multipoles 
(recall that the non-resonant phase-shifts are zero at NLO), 
the Fermi-Watson theorem in the resonant channels is violated. 
In particular, one sees that the real parts 
of the resonant multipoles  now fail to cross zero at the resonance position.
The complete NLO calculation, shown by the solid curves in the figure
includes in addition the $\pi N$-loop corrections in \Figref{NLOvertex}, 
which restore unitarity. The Fermi-Watson theorem is satisfied 
exactly in this calculation.

\begin{figure}[t]
\centerline{  \epsfxsize=9cm
  \epsffile{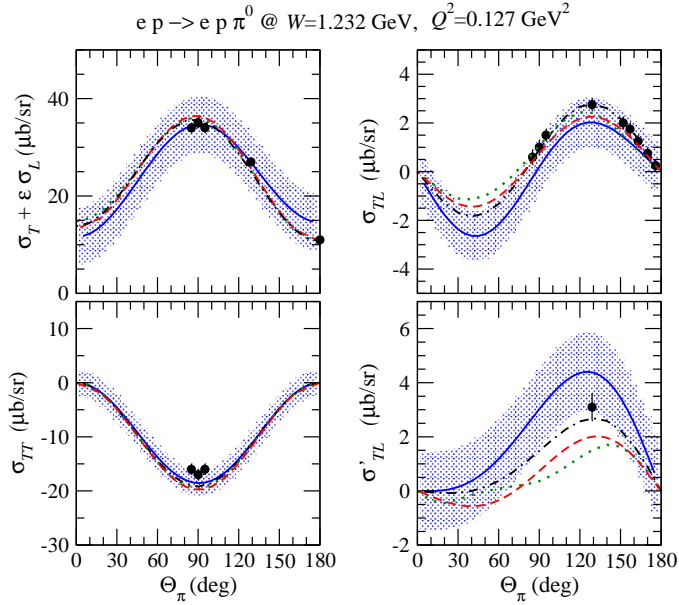}
}
\caption{The pion angular dependence
of the $\ga^\ast p \to \pi^0 p$ cross sections at
$W = 1.232$~GeV and $Q^2$ = 0.127~GeV$^2$.
Dashed-dotted (black) curves: DMT model~\cite{KY99}.
Dashed (red) curves: SL model~\cite{SL}.
Dotted (green) curves : DUO model~\cite{DUO}.
Solid (blue) curves: 
\ceft \ results \cite{Pascalutsa:2005ts,Pascalutsa:2005vq}.
The bands provide an estimate of the theoretical error for the \ceft \
calculations.
Data points are from BATES
experiments~\cite{Mertz:1999hp,Kunz:2003we,Sparveris:2004jn}.
}
\label{fig:epio_cross2}
\end{figure}
In Fig.~\ref{fig:epio_cross2},
the different virtual photon absorption cross sections
around the resonance position are displayed
at $Q^2 =0.127$~GeV$^2$,
where recent precision data are available.
We compare these data with the present 
\ceft \ calculations as well as with the results of
SL, DMT, and DUO models~\cite{SL,KY99,DUO}.

In the \ceft \ calculations, the low-energy constants
$g_M$ and $g_E$, were fixed from the resonant pion photoproduction
multipoles. Therefore, the only other low-energy constant from the
chiral Lagrangian entering the NLO calculation is $g_C$. The main
sensitivity on $g_C$ enters in $\sigma_{TL}$. A best description
of the $\si_{TL}$ data at low $Q^2$ is obtained by choosing $g_C =
-2.6$. 

From the figure one sees that
the NLO \ceft\ calculation, within its accuracy, is consistent
with the experimental data for these observables at low $Q^2$.
The dynamical models are in basic agreement with each
other and the data for the transverse cross sections. Differences
between the models do show up in the $\sigma_{TL}$ and
$\sigma_{TL'}$ cross sections which involve the longitudinal
amplitude. In particular for $\sigma_{TL'}$ the differences
reflect to a large extent how the non-resonant $S_{0+}$ multipole
is described in the models.

\section{Pion-mass dependence of the $\ga N\De$
form factors}

Since the low-energy constants $g_M$, $g_E$, and $g_C$ have been fixed, our  
calculation can provide a prediction for the $m_\pi$ dependence of the 
$\gamma N \Delta$ transition form factors. The study of the $m_\pi$-dependence 
is crucial to connect to lattice QCD results, which at present 
can only be obtained for 
larger pion masses (typically $m_\pi \gtrsim 300$ MeV). 
\begin{figure}[t,b]
\centerline{ \epsfxsize=7cm%
\epsffile{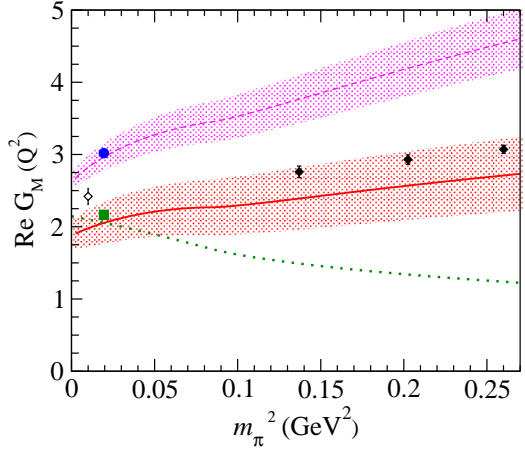}
}
\caption{
The pion mass dependence of the real part of 
the Jones-Scadron $\gamma^* N \Delta$ form factor $G_M^\ast$ for $Q^2 = 0$ and 
$Q^2$ = 0.127 GeV$^2$ in the $\chi$EFT framework.
The solid (dashed) curves are the NLO results for $Q^2 = 0.127$~GeV$^2$
($Q^2 = 0$) respectively, including
the $m_\pi$ dependence of $M_N$ and $M_\Delta$. 
The green dotted curve is 
the corresponding result for $Q^2 = 0.127$~GeV$^2$ where   
the $m_\pi$ dependence of $M_N$ and $M_\Delta$ is not included. 
The blue circle for $Q^2 = 0$ 
is a data point from MAMI~\protect\cite{Beck:1999ge}, and 
the green square for $Q^2 = 0.127$~GeV$^2$ is 
a data point from BATES~\protect\cite{Mertz:1999hp,Sparveris:2004jn}. 
The three filled black diamonds at larger $m_\pi$   
are lattice calculations~\protect\cite{Alexandrou:2004xn} for $Q^2$ values of 
0.125, 0.137, and 0.144 GeV$^2$ respectively, 
whereas the open diamond near $m_\pi \simeq 0$ represents their  
extrapolation assuming linear dependence in $m_\pi^2$. 
}
\label{fig:regmmpi}
\end{figure}

In Fig.~\ref{fig:regmmpi} we examine the $m_\pi$-dependence of the 
{\it magnetic} $\gamma N \Delta$-transition form factor $G_M^\ast$, in  
the convention of Jones and Scadron~\cite{Jones:1972ky}. 
At the physical pion mass, this form 
factor can be obtained from the imaginary part of the $M_{1+}^{3/2}$ multipole 
at $W = M_\Delta$ (where the real part is zero by Watson's theorem).  
The value of $G_M^\ast$ at $Q^2 = 0$ is 
determined by the low-energy constant $g_M$.
The $Q^2$-dependence then follows as a prediction of the NLO
result, and Fig.~\ref{fig:regmmpi} shows that
this prediction is consistent with the experimental 
value at $Q^2 = 0.127$~GeV$^2$ and physical pion mass. 

The $m_\pi$-dependence  
of $G_M^\ast$ is also completely fixed at NLO, no new parameters appear. 
In Fig.~\ref{fig:regmmpi}, the result for $G_M^\ast$ at
$Q^2 = 0.127$~GeV$^2$ is shown both when the $m_\pi$-dependence of 
the nucleon and $\Delta$ masses is included (solid line) 
and when it is not (dotted line).
Accounting for the $m_\pi$-dependence in $M_N$ and $M_\Delta$,
significantly affects $G_M^\ast$. 
The \ceft\  calculation,  
with the $m_\pi$ dependence of $M_\N$ and $M_\Delta$ included, is in
a qualitatively good agreement with the lattice data shown in the figure. 
The \ceft\ result also follows 
an approximately linear behavior in $m_\pi^2$, 
although it falls about 10 - 15 \% below the lattice data.  
This is just within the uncertainty of the NLO results.
One should also keep in mind that the present lattice simulations
are not done in full QCD, but are ``quenched'', so discrepancies
are not unexpected.

\begin{figure}[b,t]
\centerline{  \epsfxsize=7cm%
  \epsffile{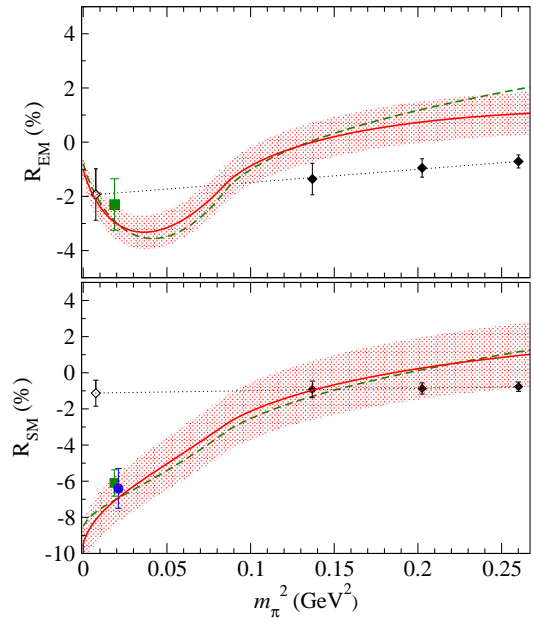} 
}
\caption{
The pion mass dependence of
 $R_{EM}$ (upper panel) and 
$R_{SM}$ (lower panel), at $Q^2=0.1$ GeV$^2$.
The blue circle is a data point from MAMI~\protect\cite{Pospischil:2000ad}, 
the green squares are data points from 
BATES~\protect\cite{Mertz:1999hp,Sparveris:2004jn}. 
The three filled black diamonds at larger $m_\pi$   
are lattice calculations~\protect\cite{Alexandrou:2004xn}, 
whereas the open diamond near $m_\pi \simeq 0$  
represents their extrapolation assuming linear dependence in $m_\pi^2$. 
Red solid curves: NLO result when accounting for the $m_\pi$ dependence in 
$M_N$ and $M_\Delta$; 
green dashed curves: NLO 
result of Ref.~\cite{Pascalutsa:2005ts}, where   
the $m_\pi$-dependence of $M_N$ and $M_\Delta$ was not accounted for.
The error bands represent the estimate of theoretical uncertainty for
the NLO calculation. 
}
\figlab{ratios}
\end{figure}
In \Figref{ratios}, we show the $m_\pi$-dependence of the ratios 
$R_{EM}$ and $R_{SM}$ and compare them to lattice QCD calculations.  
The recent state-of-the-art lattice calculations of 
$R_{EM}$ and $R_{SM}$~\cite{Alexandrou:2004xn} use a {\it linear}, 
in the quark mass ($m_q\propto m_\pi^2$), {\it extrapolation}
to the physical point,  
thus assuming that the non-analytic $m_q$-dependencies are  negligible. 
The thus obtained value for $R_{SM}$ at the physical 
$m_\pi$ value displays a large 
discrepancy with the  experimental result, as seen in \Figref{ratios}. 
Our calculation, on the other hand, shows  that the non-analytic dependencies 
are {\it not} negligible. While
at larger values of $m_\pi$, 
where the $\Delta$ is stable, the ratios display a smooth 
$m_\pi$ dependence, at $m_\pi =\De $ there is an inflection point, and 
for  $m_\pi \leq \Delta$ the non-analytic effects are crucial, 
as was also observed for the $\De$-resonance
magnetic moment~\cite{Cloet03,PV05}.

One also sees from \Figref{ratios} that, unlike the result for 
$G_M^\ast$, there is little difference between the \ceft\ 
calculations with the $m_\pi$-dependence of 
$M_N$ and $M_\Delta$ accounted for, and our earlier calculation
\cite{Pascalutsa:2005ts}, 
where the ratios were evaluated neglecting 
the $m_\pi$-dependence of the masses. 
This is easily understood, as the main effect due to the $m_\pi$-dependence 
of $M_N$ and $M_\Delta$ arises due to a common factor in the evaluation 
of the $\gamma N \Delta$ form factors, which drops out of the ratios.
One can speculate that the ``quenching'' effects 
drop out, at least partially, from the ratios as well.  
In \Figref{ratios} we also show the $m_\pi$-dependence of the $\gamma N \De$ transition ratios, 
with the theoretical error band. 
The $m_\pi$ dependence obtained here from \ceft\  clearly shows that
the lattice results for $R_{SM}$ may in fact be consistent 
with experiment.

\section{Conclusion}
\label{sec7}

We presented here an extension of the chiral perturbation theory
framework to the energy domain of $\De$(1232). In this extension
the $\De$-isobar appears as an explicit degree of freedom in the
effective chiral Lagrangian. The other low-energy degrees of freedom
appearing in this Lagrangian are pions, nucleons, and $\rho$-mesons.
The power counting depends crucially on how the $\De$-resonance
excitation energy, $\vDe=M_\De-M_N$, compares to the other scales
in the problem. In the ``$\de$-expansion'' scheme adopted here, one utilizes
the scale hierarchy, $m_\pi \ll \vDe \ll \La_{\chi SB}$. In other
words, we have an EFT with two distinct light scales. Such a scale 
hierarchy is crucial for an adequate counting
of the $\De$-resonance contributions in both the low-energy
and the resonance energy regions. The $\de$ power counting
provides a justification for ``integrating out'' the resonance
contribution at very low energies, as well as for resummation
of certain resonant contributions in the resonance region. 

We applied the $\de$-expansion to the process of
pion electroproduction. 
This is a first \ceft\ study of this reaction
in the $\Delta(1232)$-resonance region.
Our resulting next-to-leading order calculation was shown to
satisfy the electromagnetic gauge and chiral symmetries,
Lorentz-covariance, analyticity, unitarity (Watson's theorem).
The free parameters entering at this order
are the $\ga N\De$ couplings $g_M$, $g_E$, $g_C$ characterizing
the $M1$, $E2$,  $C2$ transitions, respectively. 
By comparing our NLO results with the standard 
multipole solutions (MAID and SAID) for the photoproduction multipoles
we have extracted $g_M$ = 2.97 and 
$g_E = -1.0$, corresponding to $R_{EM} = -2.2$~\%.  
The NLO \ceft\ 
result was also found to give a good description of the 
energy-dependence of most non-resonant $s$, $p$ and $d$-wave 
photoproduction multipoles in a 100 MeV window 
around the $\Delta$-resonance position. 
From the pion electroproduction cross-section $\si_{LT}$
we have extracted $g_C = -2.6$, which yields $R_{SM} \simeq -7$~\% 
near $Q^2 = 0.127$~GeV$^2$. In overall, the NLO
results are consistent with the experimental data
of the recent high-precision measurements at MAMI and BATES.  

The \ceft\ framework plays a {\it dual role}
in that it allows for
an extraction of resonance parameters from observables {\em and} 
predicts their pion-mass dependence. In this way it may provide
 a crucial connection of present lattice QCD results (obtained 
at larger than physical values of $m_\pi$) to the experiment.
We have shown here that the opening of the 
$\De\to \pi N$ decay channel at $m_\pi = M_\De-M_N$
induces a pronounced  non-analytic
behavior of the $R_{EM}$ and $R_{SM}$ ratios. 
While the linearly-extrapolated lattice QCD results 
for $R_{SM}$ are in disagreement with experimental data, the 
\ceft\ prediction of the non-analytic dependencies suggests that
these results are in fact consistent with experiment.

The presented results are systematically improvable. 
We have indicated what are the next-next-to-leading order effects,
however,  at present we could 
only estimate the theoretical uncertainty of our calculations
due to such effects. We have defined and provided a corresponding
error band on our NLO results. An actual calculation of 
N$^2$LO effects is a worthwhile topic for a future work.

\begin{theacknowledgments}
We thank Aron Bernstein and Costas Papanicolas for invitation
to this stimulating workshop and generous support.
This work is partially supported by DOE grant no.\
DE-FG02-04ER41302 and contract DE-AC05-06OR23177 under
which Jefferson Science Associates operates the Jefferson Laboratory.  
\end{theacknowledgments}

\end{document}